\documentclass[aip,
jmp,
amsmath,amssymb,preprint,jcp
reprint,%
]{revtex4-1}

\usepackage{comment}
\usepackage{hyperref}
\usepackage{placeins}
\usepackage{wrapfig}
\usepackage{lipsum}
\usepackage{float}
\usepackage{graphicx}
\graphicspath{ {./images/} }
\usepackage{subcaption}
\usepackage{pgfplotstable}
\usepackage[export]{adjustbox}
\usepackage{dcolumn}
\usepackage{bm}
\usepackage{siunitx}

\usepackage{color}
\usepackage[utf8]{inputenc}
\usepackage[T1]{fontenc}
\usepackage{mathptmx}
\usepackage{etoolbox}
\makeatletter
\def\@email#1#2{%
 \endgroup
 \patchcmd{\titleblock@produce}
  {\frontmatter@RRAPformat}
  {\frontmatter@RRAPformat{\produce@RRAP{*#1\href{mailto:#2}{#2}}}\frontmatter@RRAPformat}
  {}{}
}%
\makeatother

\begin{document}

\preprint{~}

\title[The surface tension of Martini 3 water mixtures]{The surface tension of Martini 3 water mixtures}

\author{Lorenzo Iannetti}
\affiliation{Dipartimento di Ingegneria Meccanica ed Aerospaziale, Sapienza Università di Roma, Via Eudossiana 18, 00184 Roma, Italia
}
\author{Sonia Cambiaso}
\affiliation{Dipartimento di Fisica, Università of Genova, via Dodecaneso 33, 16146 Genova, Italia}
\author{Fabio Rasera}
\affiliation{Dipartimento di Ingegneria Meccanica ed Aerospaziale, Sapienza Università di Roma, Via Eudossiana 18, 00184 Roma, Italia
}

\author{Alberto Giacomello}
\affiliation{Dipartimento di Ingegneria Meccanica ed Aerospaziale, Sapienza Università di Roma, Via Eudossiana 18, 00184 Roma, Italia 
}
\author{Giulia Rossi}
\affiliation{Dipartimento di Fisica, Università of Genova, via Dodecaneso 33, 16146 Genova, Italia}
\author{Davide Bochicchio}
\affiliation{Dipartimento di Fisica, Università of Genova, via Dodecaneso 33, 16146 Genova, Italia} 

\author{Antonio Tinti$^\ast$}
\email{antonio.tinti@uniroma1.it}
\affiliation{Dipartimento di Ingegneria Meccanica ed Aerospaziale, Sapienza Università di Roma, Via Eudossiana 18, 00184 Roma, Italia
}

\date{\today}

\maketitle

\section{\label{sec:abstract}Abstract}

The Martini model, a coarse-grained forcefield for biomolecular simulations, has experienced a vast increase in popularity in the past decade. Its building-block approach balances computational efficiency with high chemical specificity, enabling the simulation of various organic and inorganic molecules.
The modeling of coarse-grained beads as Lennard-Jones particles poses challenges for the accurate reproduction of liquid-vapour interfacial properties, which are crucial in various applications, especially in the case of water. The latest version of the forcefield introduces refined interaction parameters for water beads, tackling the well-known artefact of Martini water freezing at room temperature. Additionally, multiple sizes of water beads are available for simulating the solvation of small cavities, including the smallest  pockets of proteins. This work focuses on studying the interfacial properties of Martini water, including surface tension, surface thickness, and bulk densities for the liquid and vapour phases. Employing the test-area method, we systematically compute the liquid-vapour surface tension across various combinations of water bead sizes and for temperatures in the range from 300 to 350 K. Our findings provide  a comprehensive characterization of Martini 3.0 water intefacial properties. These findings are of interest to the Martini community as they allow users to account for the low interfacial tension of Martini water by properly adjusting observables computed via coarse-grained simulations (e.g., capillary forces)  to allow for accurate matching against all-atom or experimental results. Surface tension data are also interpreted in terms of local enrichment of the various mixture components at the liquid-vapour interface by means of Gibbs' adsorption formalism.
Finally, the critical scaling of the Martini surface tension with temperature is reported to be consistent with the critical exponent of the 3D Ising universality class, a result of general interest for Lennard-Jones mixtures.

\section{\label{sec:intro}Introduction}

Coarse-grained (CG) forcefields represent nowadays an invaluable tool for molecular dynamics (MD) simulations. By lumping several atoms into single interaction centers, called beads, they reduce the number of degrees of freedom of the system, allowing the simulation to tackle spatial and temporal scales that would not be accessible to atomistic simulations. Among the available CG models, Martini represents a popular choice for biophysical simulations. The most recent version of the Martini forcefield, Martini 3\cite{souza21_martini3}, hinges on a flexible mapping scheme in which two, three, or four heavy atoms are represented, together with the hydrogens they are bound to, by single interaction centers. The extensive availability of bead types, with different levels of polarity and the possibility to further take into account the beads'  hydrogen-bonding capability and polarizability, allows the forcefield to retain a high chemical specificity. The building-block approach of Martini, in which the different single beads' properties, and more specifically their oil-water partitioning, are targeted to experimental data, makes Martini useful to simulate an enormous variety of biological and material science systems with many degrees of freedom and complex collective behaviours, such as biomolecules\cite{chakraborty_glicans,uusitalo_dna,monticelli_proteins}, polymers\cite{rossi_ps,grunewald_peg,alessandri_p3ht,panizon_pp}, lipids\cite{marrink_lipids,baoukina_lipids,wassenaar_insane}, and nanoparticles in solution or interacting with the biological environment\cite{salassi_NP,lavagna_NP,das_NP,simonelli2019_NP}. Both in the biological and in the material science realm, Martini is often used for the simulation of liquid-liquid interfaces, such as water-surfactant \cite{anogiannakis_surfactant,rossi_surfactant,illa_surfactant}, polymer coating-water \cite{uttarwar_coatings,rossi_coatings,verma_coatings}, and surface-solvent interfaces\cite{perrin18_surface,schulze_AuSurf,wu_graphene}. 
Non-bonded interactions in Martini are described by a 12-6 Lennard-Jones potential, which is a relatively hard interaction potential, known for limiting the range of the liquid phase in favor of the solid one. This limitation has suggested not to use free energies of vapourization and solvation, nor gas-liquid surface tensions \cite{souza21_martini3} as targets for the parameterization of the Martini CG model, as reproducing these targets would exacerbate the tendency of the simple LJ fluids to make the transition to the solid state at unrealistically high temperatures. The Martini forcefield thus uses experimental free energies of partitioning as the main target for the parameterization of non-bonded interactions and has the well-known limitation of a scarce reproduction of the energetics of vapourization/solvation and of liquid/vapor surface tensions. This latter limitation may impact the reliability of the forcefield in specific applications requiring the simulation of liquid/vapor interfaces, but the problem can be mitigated by an \textit{ad-hoc} adjustment if the surface tension, $\gamma$, of the Martini model is known. The case of water is obviously the most relevant one. Accurate modeling of surface tension is crucial to use Martini for a vast range of wetting phenomena \cite{rauscher2008}, e.g., evapouration in confinement \cite{guillemot2012_activated, remsing, tinti2023}, water intrusion/extrusion processes in hydrophobic nanoporous materials \cite{tinti17,ledonne,cambiaso-rasera_grafting}, liquid-vapour phase change phenomena\cite{cho2016_nanoengineered}, contact line problems \cite{ren2007}, superhydrophobic surfaces \cite{kusumaatmaja2008,koishi2009,amabili2017,silvestrini2021}, and contact angle hysteresis at the nanoscale\cite{chen2022_CAhyst,giacomello2016}. 

Surface tension arises from the anisotropy of molecules at an interface. The high surface tension of water ($\gamma$=72~mN/m) is due to the presence of hydrogen bonds, causing strong electrostatic interactions\cite{martini_polarizable}; these interactions are not explicitly taken into account in Martini. As a result, for the regular water beads of the second release of the Martini forcefield \cite{marrink2007martini} the surface tension at T=293 K was measured to be $\gamma=45$~mN/m at $T=293$~K. The polarizable Martini water model further reduced the surface tension to $\gamma=30.5$~mN/m, at $T=300$~K\cite{yesylevskyy_polarizable}. 

In the latest release of the Martini forcefield, the description of water has substantially changed. Three different beads have been introduced: TW, SW, and W, representing  two, three, and four water molecules, respectively. Their mutual miscibility has been reported to be nearly ideal\cite{souza21_martini3}.  Water self-interactions have been reduced, to avoid water freezing at room temperature \cite{marrink2007martini, bruininks19_practical}. The possibility to tune the size of water molecules allows to better describe situations in which water is expected to enter small cavities, such as small protein pockets or the pores of microporous materials, while, at the same time, it can be exploited to avoid unphysical water freezing in the presence of ordered interfaces \cite{perrin18_surface,salassi_NP}. 

Several methods exist to compute the surface tension in simulations, including force integration, equilibrium fluctuation, non-equilibrium, and density gradient methods\cite{popinet2018numerical,navascues1979_liquid}. Some methods, such as the Kirkwood and Buff method\cite{KBmethod}, rely on the mechanical definition of surface tension based on the computation of the pressure tensor, while others, including the test area method\cite{gloor2005,vega2007}, are based on the thermodynamic definition. Despite its apparent simplicity, computation of surface tension in simulation can still be challenging as in real world scenarios results can be affected by finite-size effects, truncation effects, and subtleties associated to the different mechanical and thermodynamic definitions of the surface tension \cite{trokhymchuk1999, martini_polarizable,ghoufi2016computer}. 
In this work, we performed molecular dynamics simulations to systematically compute, using the test area method\cite{gloor2005}, the liquid-vapour surface tension of TW, SW, and W Martini water and of their mixtures, to allow for proper adjustment to the experimental surface tension of water whenever needed. Results show how the presence of SW and TW beads affects the interfacial tension and its universal scaling with temperature. 
   
\section{\label{sec:methods}Methods}
Water beads were modeled using the Martini 3 forcefield, which prescribes implementing the LJ interactions with a cutoff of 1.1 nm. The L-J parameters for W, SW, and TW beads are reported in Tab.~\ref{tab:martini3} 

\begin{table}[h]
\centering
\begin{tabular}{l@{\hspace{7pt}}|@{\hspace{7pt}}c@{\hspace{7pt}}c@{\hspace{7pt}}c@{\hspace{7pt}}l}
\cline{1-4}
bead & W & SW & TW &  \\ \cline{1-4}
$\epsilon$ [kJ/mol] &  4.65 &  3.20 & 2.80 &  \\ 
$\sigma$ [nm] & 0.44  & 0.41  & 0.35  &  \\ 
cutoff [nm] & 1.1 (2.5$\sigma$) &  1.1 ($\sim 2.683 \sigma$) & 1.1 ($\sim 3.143 \sigma$) & \\ 
mass [g/mol] & 72 & 54 & 36 \\ \cline{1-4}
\end{tabular}
\caption{Martini 3 interaction parameters for the different water beads.}
\label{tab:martini3}
\end{table}

\begin{figure}[H]
    \centering
    \includegraphics[width=\textwidth]{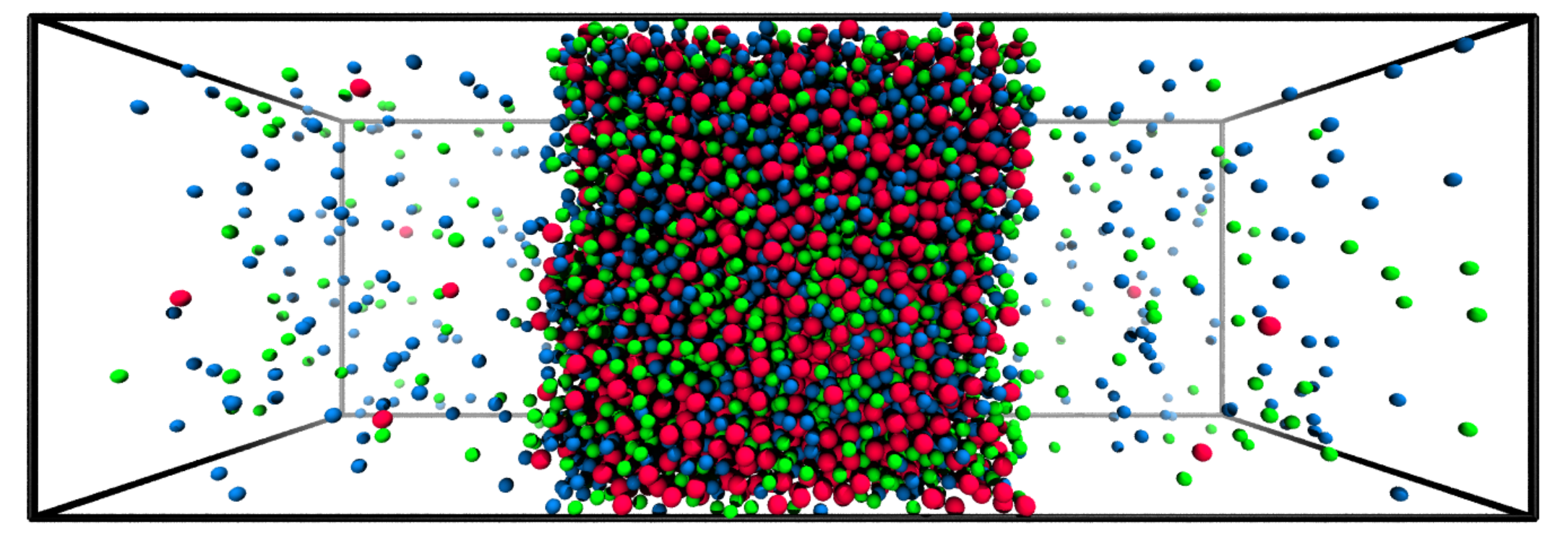}
    \captionsetup{name=Fig, font=footnotesize, justification=raggedright, singlelinecheck=false}
    \caption{MD snapshot depicting one of our systems after a 2000 ps long equilibration that resulted in liquid-vapour phase separation; this specific mixture comprises all the three bead types in equal proportion. W beads are depicted in red, SW beads in green and TW beads in blue.}
    \label{fig:simbox}
\end{figure}

As customary in MARTINI simulations, a timestep of 20 fs was applied and PBC were used along all the directions during all the steps of the simulations. The surface tension was computed using the test area method for several Martini waters compositions spanning all possible combinations of beads of the three types (W, SW, TW) in 10\% increments. The original system, containing 11000 water beads in the specified nominal molar ratios within a cubic box of 1000 $\text{nm}^3$, was initially equilibrated in NPT conditions (constant number of particles, pressure, and temperature)  for 2000 ps at 1 atm and 300 K to achieve the target density. Two empty boxes of equal dimensions were then added along the z axis to allow for phase separation and the formation of a planar vapour-liquid interface along the xy plane. Subsequently the system was newly equilibrated in the NVT ensemble (at constant number of particles, volume, and temperature) for another 2000 ps at the same temperature to stabilize the liquid-vapour interface. Fig.~\ref{fig:simbox} shows an example of how the final simulated compositions appear after the equilibration procedure highlighted above.  

To avoid large fluctuations of the interface the total momentum of the system was set to zero at each timestep. 

A production run of 1.5 $\mu$s for each composition was launched to compute the surface tension, following the test area method introduced by Gloor et al.\cite{gloor2005}. In the test area method the axes parallel to the \emph{l-v} interface are perturbed by a scaling factor $\left(1+\delta\right)^{\frac{1}{2}}$ while the remaining axis is scaled by factor $\left(1+\delta\right)^{-1}$ in order to conserve the total volume. The potential energy variation $\Delta U$ in the system is then computed for each perturbation of the \emph{l-v} interface and the surface tension is obtained using the equation 

        \begin{equation}
            \gamma = \left(\frac{\partial F}{\partial A}\right)_{N, T, V} = -\frac{k_BT}{\Delta A}\ln\left<\exp\left(\frac{\Delta U}{k_BT}\right)\right>
        \end{equation}
        
 where $F$ is the Helmholtz free energy, $k_B$ the Boltzmann constant, $T$ the absolute temperature, $\Delta A = L_xL_y\delta$ the infinitesimal variation of the interface area, and the brackets $\left<\cdot\right>$ denote an average over the equilibrium configurations.  $\delta$ must be chosen sufficiently small such that the value of $\gamma$ does not depend on it (see Fig.~\ref{fig:test_area} in Appendix~\ref{appendix:comparison} and \ref{appendix:virial}). In our simulations a factor $\delta = 10^{-4}$ was chosen and the system was perturbed every 300 ps to avoid correlated configurations.
 The same procedure was repeated at higher temperatures for all the compositions, between 300 and 350 K, with 10 K intervals.

\section{\label{sec:results}Results}

\subsection{Phase separation and density profiles}
\begin{figure}[h]
            \centering
            \includegraphics[width=\textwidth]{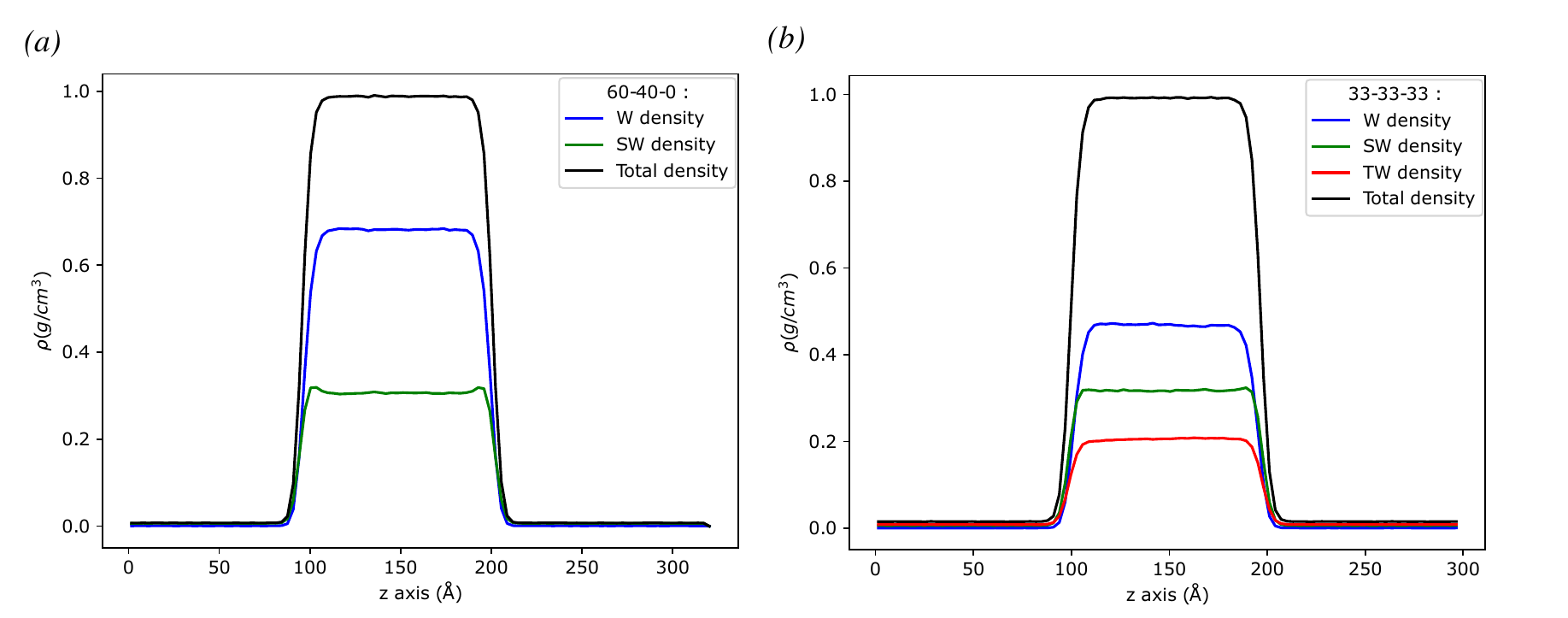}
            \captionsetup{name=Fig, font=footnotesize, justification=raggedright, singlelinecheck=false}
            \caption{Mass density profiles displaying liquid-vapour phase separation and the appearance of two planar interfaces for (a) 60\% regular W beads and 40\% SW beads, (b) 33\% for all three bead types.}
            \label{fig:density}
        \end{figure}

From the trajectories, we computed the average density profiles along the z axis for all the mixtures, see Fig.~\ref{fig:density}. 
In order to fully characterize the \emph{l-v} interface, the density profiles were fitted in the vicinity of the \emph{l-v} interface using the well known expression \cite{rowlinson}:

        \begin{equation}
            \rho (z) = \frac{\left(\rho_l-\rho_v\right)}{2}\left(\tanh\left(\frac{z_0-z}{\lambda}\right)+1\right) + \rho_v
        \end{equation}
        
where $z_0$ is the position of the Gibbs dividing interface, $\rho_l$ and $\rho_v$ are  the liquid and vapour bulk densities, respectively, and $\lambda$ is the thickness of the interface (Fig.~\ref{fig:density}a,b). Using this expression the bulk densities $\rho_l, \rho_v$, the interface thickness $\lambda$, and the position of the Gibbs dividing interface $z_0$ were obtained as fitting parameters.   

 In Tab.~\ref{tab:densities}, we report the measured bulk liquid and vapour densities for the pure Martini 3 water species at 300 K. The values of bulk liquid density were found to be very close to the experimental one. The vapour bulk density, instead, varies significantly across the different kinds of water depending on their size, with TW displaying the most volatile behaviour. The interface thickness at 300 K was in the range of 0.3 nm and increased with temperature up to values of the order of 1 nm at 350 K, depending on the mixture composition. The complete data for $\rho_l$, $\rho_v$, and $\lambda$ are reported in the Supplementary Materials. 

\begin{table}[h]
\centering
\begin{tabular}
{c@{\hspace{7pt}}|@{\hspace{7pt}}c@{\hspace{15pt}}c@{\hspace{15pt}}c}
\hline
& W & SW & TW \\ \hline
$\rho_l(g/cm^3)$ & 0.9843 & 0.9812 & 0.9954  \\ 
$\rho_v(g/cm^3)$ & 0.0007 & 0.0169 & 0.0330 \\ \hline
\end{tabular}
\captionsetup{name=TAB, font=footnotesize, justification=raggedright, singlelinecheck=false}
\caption{Liquid and vapour bulk mass densities as measured for pure systems of the three bead types at 300 K at coexistence conditions.}
\label{tab:densities}
\end{table} 

After the equilibration procedure previously reported, the actual equilibrium concentrations of the various species in the liquid phase were found to deviate slightly from the nominal molar concentrations imposed in the initial system. 
For instance, in Fig.~\ref{fig:density}b, which reports results for the case of an equal nominal concentration of  W and TW beads, the ratio between the liquid bulk mass densities should be equal to two since the mass of W beads, modeling four water molecules, is twice the mass of TW beads, that models only two water molecules (see Tab.~\ref{tab:martini3}). However, the actual ratio is slightly higher than the  nominal value due to the fact that smaller beads tend to be more volatile and as such they have a higher preference for the vapour phase. Tab.~\ref{tab:equilfraction} in the Appendix~\ref{appendix:mol_fractions} reports the equilibrium molar fractions of beads in the liquid and vapour bulks for binary mixtures. 
In the reminder of the paper, in order to simplify exposition of the results, nominal concentration values will be used to refer to the different simulated conditions, yet
the molar fractions of beads used in calculations will refer to the actual equilibrium values that were measured in the liquid bulk .

\subsection{Surface tension measurements}
\begin{figure}[hb!]
        \centering
        \includegraphics[width=\textwidth]{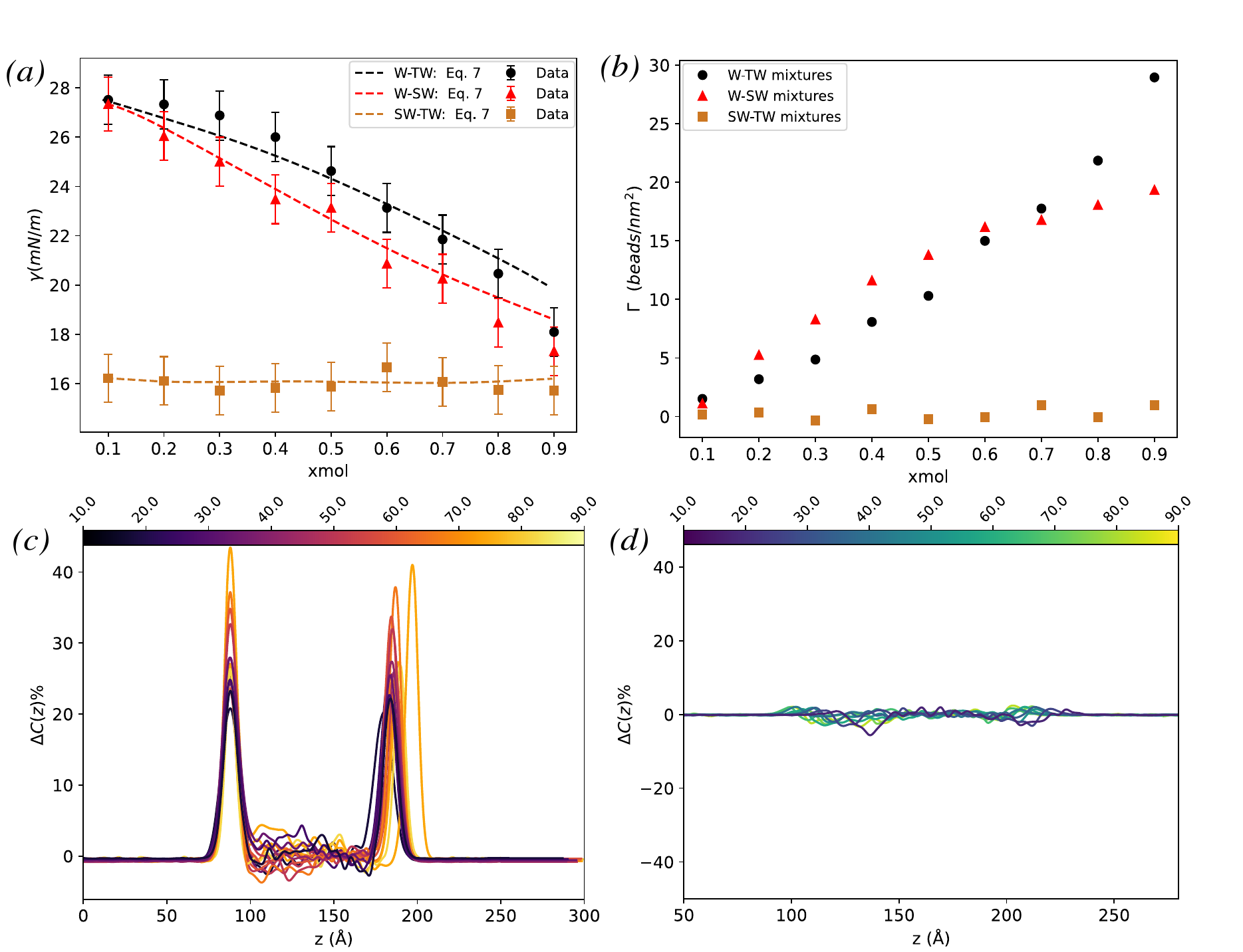}
        \captionsetup{name=Fig, font=footnotesize, justification=raggedright, singlelinecheck=false}
        \caption{\footnotesize (a) Comparison between surface tension obtained from simulations and theoretical predictions from eq. \eqref{eqn:gibbs} and \eqref{eq:ideal_mixtures}; the x axis represents the nominal molar fraction of TW beads for W-TW and SW-TW mixtures, and of SW beads in the case W-SW mixtures. Statistical error bars are smaller than 1 mN/m for all the computed values of surface tension. (b) Relative adsorption of W-TW, W-SW, SW-TW binary mixtures; the x axis values represent molar fractions of the second component in the mixture (TW for W-TW and SW-TW, and SW for W-SW). It is possible to notice how SW-TW mixtures do not seem to show segregation or adsorption at the surface. (c) Surface segregation profiles for W-TW and W-SW binary mixtures expressed as excess percentage of beads with respect to the bulk: $\Delta C(z) \% = \frac{\rho(z)-\rho_{liq}}{\rho_{liq}-\rho_{vap}} \cdot 100$ ; colors indicate the nominal percentage of W beads in the mixture. (d) Surface segregation profiles for SW-TW binary mixtures as in (c), with colors indicating the nominal percentage of SW beads.}   
        \label{fig:binary_mistures}
    \end{figure}

We present here our results concerning the surface tension of MARTINI water bead mixtures. For the sake of simplicity, we  start by introducing results for pure systems and binary mixtures of the available water MARTINI 3 beads (W-SW, W-TW, and SW-TW) at different concentrations (Fig.~\ref{fig:binary_mistures}).
At 300 K the surface tension for pure systems of TW and SW beads are measured to be 15.94 and 16.26 mN/m, respectively. These values, which are very similar, are much smaller than the 28.63 mN/m value measured for the pure W case. This consideration can be extended to the case of 
the case of binary mixtures,  for which the interfacial tension is observed to be mostly influenced by the fraction of regular water W in the composition, irrespective of the detailed composition of lighter beads that is introduced in the mixture (see Fig.~\ref{fig:binary_mistures}a). The trends observed for the W-SW and W-TW mixtures effectively reflect an almost linear dependence of the surface tension of the mixtures with the nominal molar fraction of W beads, with  the two smaller types resulting effectively undistiguishable in terms of their effect on the interfacial tension.

From a rigorous standpoint the variations of surface tension with chemical composition can be rationalized by invoking the Gibbs adsorption formalism \cite{rowlinson}. Within this framework, variations in the interfacial tension as a function of the composition are attributed 
 to the differential accumulation of species at the interface as measured by the interfacial adsorption $\Gamma_{i}$  \cite{douglas_fluids, telo1981}.
 More specifically it is possible to write:
 \begin{equation}
     d\gamma = -\sum_{i=1}^{N_{s}}\Gamma_{i}d\mu_{i} \quad  \text{ with } \quad \Gamma_i = \frac{n_{i}^{Tot}-n_{i}^{\alpha}-n_{i}^{\beta}}{A} ,
     \label{eqn:gibbs}
 \end{equation}
where $n_{i}^{Tot}$ is the total number of beads of the $i$-th species, $n_{i}^{\alpha}$ and $n_{i}^{\beta}$ are the numbers of beads of the same species in the bulk phases $\alpha$ and $\beta$ and A is the area of liquid-vapour (\emph{l-v}) interface. In a binary mixture, if we define the Gibbs dividing interface such that the adsorption of the first component $\Gamma_1$ is zero and we focus on the adsorption of the second component with respect to the first one we can substitute $\Gamma_i$ with the relative adsorption $\Gamma_{21}$ \cite{Winkelmann1999}:

 \begin{equation}
     \Gamma_{21}(x_{2}) = -\alpha_{2}\int_{-\infty}^{+\infty}\Delta C(z, x_{2}) dz \, 
 \end{equation}
where $\Delta C\left(z\right)$ denotes the symmetrized surface segregation, a function of the position $z$ that is expressed in terms of the density profile as \cite{Winkelmann1999,telo1981}:
 \begin{equation}
     \Delta C(z) = \frac{\rho_{1}(z)-\rho_{liq,1}}{\alpha_{1}} - \frac{\rho_{2}(z)-\rho_{liq,2}}{\alpha_{2}}.
 \end{equation}
 
Here symbols $\rho_1(z)$ and $\rho_2(z)$ are used to denote the number densities as a function of the coordinate $z$ that is orthogonal to the liquid-vapour interfaces, whereas the symbol $\alpha_i$ was used to denote  the symmetrized concentrations the i-th species:
 
 \begin{equation}
     \alpha_{i} = \frac{\rho_{liq,i}-\rho_{vap,i}}{(\rho_{liq,1}+\rho_{liq,2})-(\rho_{vap,1}-\rho_{vap,2})}. 
 \end{equation} 
Bulk liquid and vapour number densities of the i-th species enter this expression as $\rho_{liq,i}$ and $\rho_{vap,i}$. By making use of this notation, the Gibbs isotherm for our binary mixture can be written more simply as:
\begin{equation}
    d\gamma\left(x_2\right) = -\Gamma_{21}(x_2)d\mu\left(x_2\right)
\end{equation}
which allows to make explicit the functional dependence on the molar fraction of the second component $x_2$ in the liquid phase. The chemical potential of the second species in  the equilibrium vapour can now be approximated by its ideal expression, which solely depends on the molar fraction of the second species in the vapour phase). The Gibbs adsorption isotherm now reads :
\begin{equation}
    \Gamma = -\left(\frac{\partial \gamma}{\partial x_2}\right) \left ( \frac{\partial x_{2}}{\partial x_{2v}} \right ) \frac{x_{2v}}{RT}
    \label{eq:ideal_mixtures}
\end{equation}

Where $x_{2v}$ is the equilibrium  molar fraction of the second component in vapour phase. This differential equation can finally be integrated to yield the theoretical prediction of the surface tension for ideal binary mixtures:
\begin{equation}
    \gamma\left(x_2\right) = -RT\int_{0}^{x_{2}}\frac{\Gamma_{21}\left(x_2\right)}{x_{2v}} \left ( \frac{\partial x_{2v}}{\partial x_2} \right ) ~ \text{d}x_2 + \gamma(x_2=0) 
    \label{eqn:gamma_theory}
\end{equation}
Where  $\gamma(x_2=0)$ denotes surface tension for the pure system of the component 1, that was obtained from simulation, in absence of a second component.
Relative adsorption of W-TW, W-SW, SW-TW binary mixtures are shown in Fig.~\ref{fig:binary_mistures}b. In Fig.~\ref{fig:binary_mistures}c,  representative segregation profiles $\Delta C\left(z\right)$ are shown for binary W-TW and W-SW binary mixtures computed at 300 K. The two peaks at the sides of the liquid bulk reveal that the TW and SW beads tend to be more concentrated at the \emph{l-v} interfaces as compared to the bulk, enriching them, while the concentration of W beads increases in the bulk of the liquid slab. On the contrary, the SW-TW mixtures are characterized by an almost perfect mixing, resulting in a nearly flat segregation profile for all possible combinations of these bead types (see Fig.~\ref{fig:binary_mistures}d).
In Fig.~\ref{fig:binary_mistures}a, we compare the values of the surface tension simulated with the test area method against those obtained by integrating eq.~\eqref{eqn:gamma_theory}. 
The predictions for binary mixtures match the independently simulated values and slightly deviate only at high concentrations of the species 2, as expected for ideal mixtures.
In conclusion, a simple linear interpolation of surface tension values could effectively describe the values of surface tension for the mixtures at hand, yet the Gibbs adsorption isotherm is necessary to accurately predict the dependence on composition by accounting for the segregation of smaller beads at the interfaces.  Furthermore Gibbs formalism is easily applied, as the use of the ideal gas approximation for the quite dilute vapour bulks was found to yield negligible deviations from the independently measured surface tension values.

\begin{figure}[h]
        \centering
        \includegraphics[width=0.8\textwidth, center]{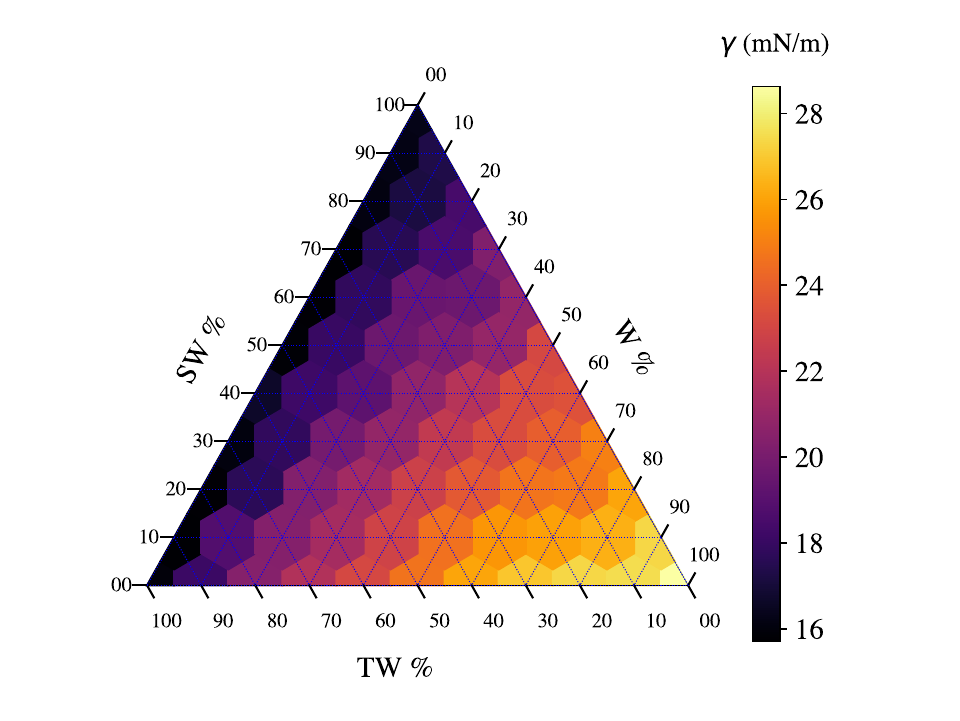}
        \captionsetup{name=Fig, font=footnotesize, singlelinecheck=false}
        \caption{Ternary diagram of the surface tension for different mixtures of Martini water beads computed at 300 K. Axes show compositions for each bead type with percentages from 0 to 100\%.}
        \label{fig:ternary}
    \end{figure}

As binary mixture cases have been thorougly discussed we can now venture into discussing simulation results for the interfacial tension of ternary systems with arbitrary concentrations of the three bead types. Test area interfacial tensions for these systems are reported in Fig.~\ref{fig:ternary}. 
Consistently with previous observations (Fig.~\ref{fig:binary_mistures}) the surface tension of the mixtures is found to almost exclusively depend on the fraction of W regardless of whether TW or SW beads constitute the remaining components of the mixture. Its value increases monotonically with the  percentage of W beads starting from 15.94 mN/m for pure TW to 28.63 mN/m for pure W systems. Finite size effects, whose effects have been considered in our measurements, are discussed in Appendix~\ref{appendix:size_effects}.

    \begin{figure}[h]
        \includegraphics[width=\textwidth]{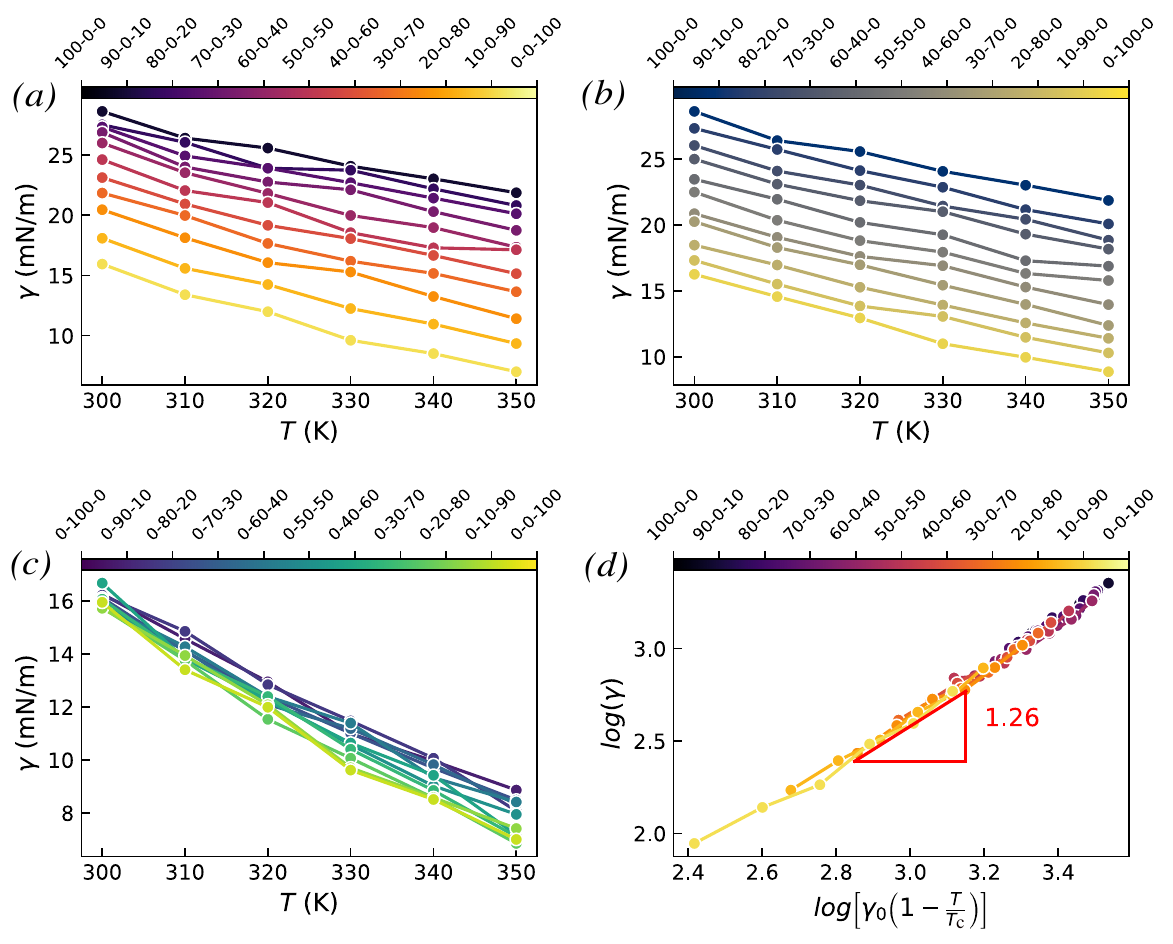}  
        \captionsetup{name=Fig, font=footnotesize, singlelinecheck=false}  
        \caption{Surface tension computed for temperatures between 300 and 350 K for a set of (a) W-TW bead mixtures, (b) W-SW mixtures, and (c) SW-TW mixtures. The set of three numbers in the upper side legend indicates the percentages of W, SW, and TW beads. d) Log-log scaling limit collapse of the surface tension for a set of W-TW mixtures. The slope of the curve appears to be in excellent agreement with the critical exponent $\mu = (d-1)\nu=2\nu=2\cdot 0.62999(5)\sim 1.26$ calculated for the 3D Ising universality class.}     
        \label{fig:gamma_T}
    \end{figure} 

\subsection{Temperature dependence of surface tension}
We now introduce the results obtained by repeating the interfacial tension measurements at different temperatures in the range $300-350$K.
Figures~\ref{fig:gamma_T}a-c report the detailed temperature dependence of surface tension for some representative binary mixtures.  All the values of surface tension computed at different temperatures are available in the Tables in Appendix~\ref{appendix:T_data}. Fig.~\ref{fig:gamma_T}c  demonstrates that $\gamma$ does not significantly change with the composition of TW-SW mixtures, similar to what was shown in Fig.~\ref{fig:binary_mistures} at 300K, although SW tends to have slightly higher surface tension at 350K.  

\begin{figure}[h]
 \centering
 \includegraphics[width=\textwidth]{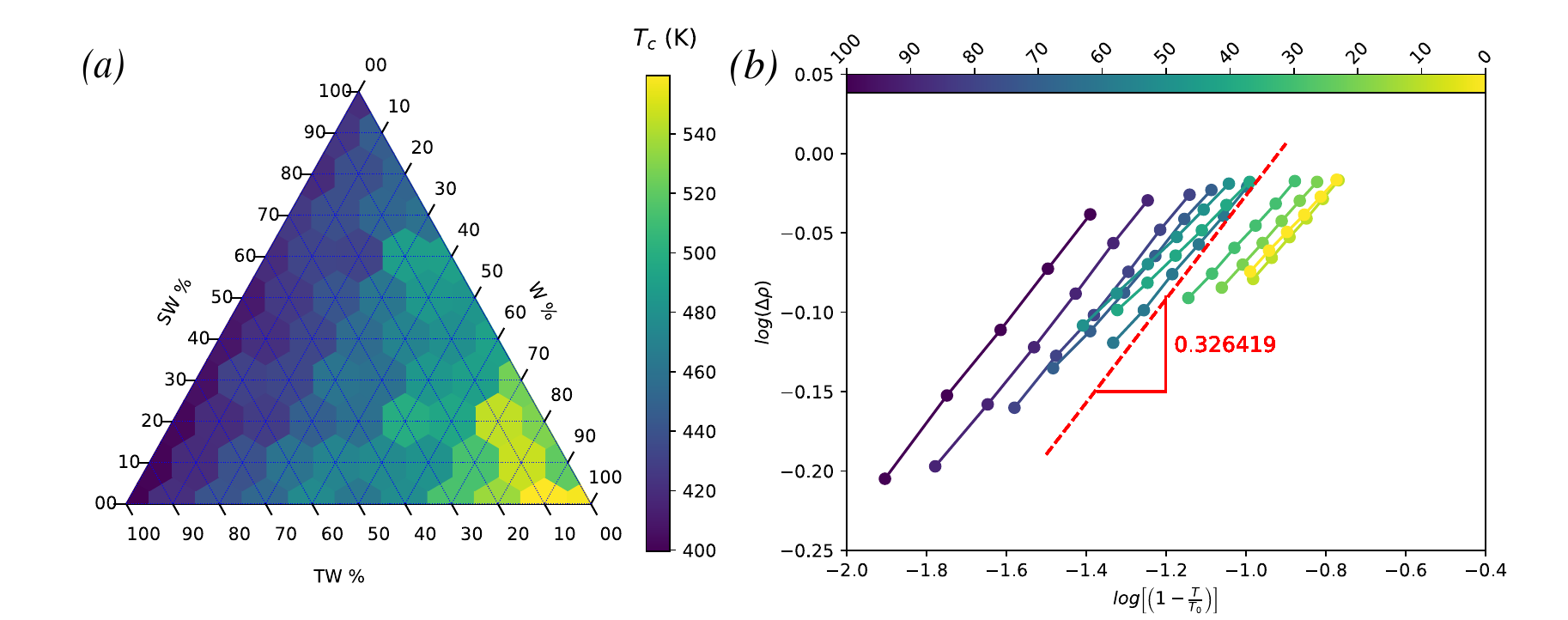}
\captionsetup{name=Fig, font=footnotesize, singlelinecheck=false}     
\caption{(a) Ternary diagram reporting critical temperatures of Martini water mixtures $T_c$ as a function of compositions. Critical temperatures were determined as the best fit for the critical scaling of the surface tension in Fig~\ref{fig:gamma_T}. (b) Log-log plot of $\Delta \rho =\rho_l-\rho_v$ as a function of $T$ for W-TW binary mixtures. Color denotes the percentage of W bead types. The scaling of the $\Delta \rho (T)$ order parameter for the various mixtures is consistent with the $\beta=\text{0.326419(3)}$ exponent which was determined for the 3D Ising universality class. The dashed red line, to be used as a guide to the eye, indicates the slope associated to this critical exponent.}
\label{fig:ordpar}
\end{figure}

The value of $\gamma$ is found to decrease with temperature in agreement with the theoretical expectations\cite{widom1972}. In particular, we could test the classic scaling of the interfacial tension with temperature that is predicted when approaching the critical temperature $T_c$ \cite{widom1972}:

    \begin{equation}
        \gamma\left(T\right) = \gamma_0\left(1 - \frac{T}{T_c}\right)^{(d-1)\nu}
        \label{eqn:gamma_scaling_law}
    \end{equation} 
    
where $d$ denotes the system's dimensionality, and $\nu$ is the critical exponent for the correlation length.
Despite the fact that, from a rigorous standpoint, this expression is expected to hold in the vicinity of the critical points of the various mixture, Figure~\ref{fig:gamma_T}d shows that,  by fitting the critical temperature $T_c$  for each composition, and by vertically shifting the curves by means of fitted $\gamma_0$ values, the interfacial tensions collapse on the  master scaling curve that is  predicted by eq.~\eqref{scaling_law} when plugging the  exponent  $\mu=(d-1)\nu=1.26$. This value appears to be remarkably close to the critical exponent $\mu=(d-1)\nu=2\nu=2 \cdot 0.62999(5)\approx 1.26$ which was calculated for the 3D Ising universality class using conformal bootstrap techniques \cite{precision}. This finding agrees with earlier investigations that indicate the 3D Ising universality class as the main candidate to represent the critical behaviour of the 3D Lennard-Jones fluid\cite{watanabe2012, panagiotopoulos1994, smit1992}. Furthermore, our results confirm the validity of such a scaling for Lennard-Jones mixtures, including Martini water bead mixtures. The values of $T_c$ resulting from the fit has been found to increase proportionally with the percentage of W beads from $\sim 400$ K at 0\% to $\sim 550$ K at 100\%, as reported in Fig.~\ref{fig:ordpar}a. 

At last, as a further consistency check we were able to make use of the fitted $T_c$ values in order to test the critical scaling of the the $\Delta \rho$ order parameter which is famously characterized by the functional dependence:
\begin{equation}
\Delta \rho = \rho_{liq} - \rho_{vap} = \rho_{0}\left(1-\frac{T}{T_c}\right)^\beta.
\label{eqn:density_scaling_law}
\end{equation}
Temperature scaling of the order parameter $\Delta \rho$, which is reported in Fig.\ref{fig:ordpar}b, provides further indication that the behaviour of our mixtures is compatible with the universality class of the 3D Ising model, that is characterized by a critical exponent $\beta= 0.326419$.

\section{\label{sec:conclusions}Conclusions}
    This work investigates the properties of the liquid-vapour interface of the newly-introduced coarse-grained Martini 3 water mixtures, with a particular focus on the liquid-vapour surface tension at biologically relevant temperatures ($300$ to $350$ K). 
    By examining all possible Martini water mixtures with different proportions of water bead sizes (W, SW, TW), we constructed a ternary diagram of the surface tension as a function of mixture composition.
    Additionally, we determined the bulk densities of vapour and liquid phases and identified the critical temperatures for each mixture.
    Our results revealed that the surface tension of Martini water effectively increases almost linearly with the percentage of W beads, irrespective of the detailed fraction of the smaller SW or TW particles. Surface tension was observed to vary in the range $\left [ 15.94-28.6  \right ]$  mN/m at $300$~K. This trend is confirmed also for ternary mixtures and at different temperatures.  
    While the surface tension of Martini water at all compositions is certainly lower than the experimental counterpart, knowing its precise value may allow material scientists to rescale capillary observables in order to obtain experimentally relevant estimates. For example, in order to estimate the force acting on a triple contact line in the presence of defects, one may multiply the value computed via coarse-grained simulation value by an "adjustment factor" given by the ratio $\gamma_\mathrm{exp}/\gamma_\mathrm{CGMD}$. Furthermore, the temperature scaling of the Martini surface tension and of the order parameter $\Delta \rho$ is found to be consistent with the critical exponents of the 3D Ising universality class, a result of more general interest for Lennard-Jones mixtures.

\section{Appendix}
\subsection{Comparison between test-area and virial methods}
\label{appendix:comparison}

\begin{figure}[H]
    \centering
    \begin{minipage}{.5\textwidth}
        \centering
        \includegraphics[width=\linewidth]{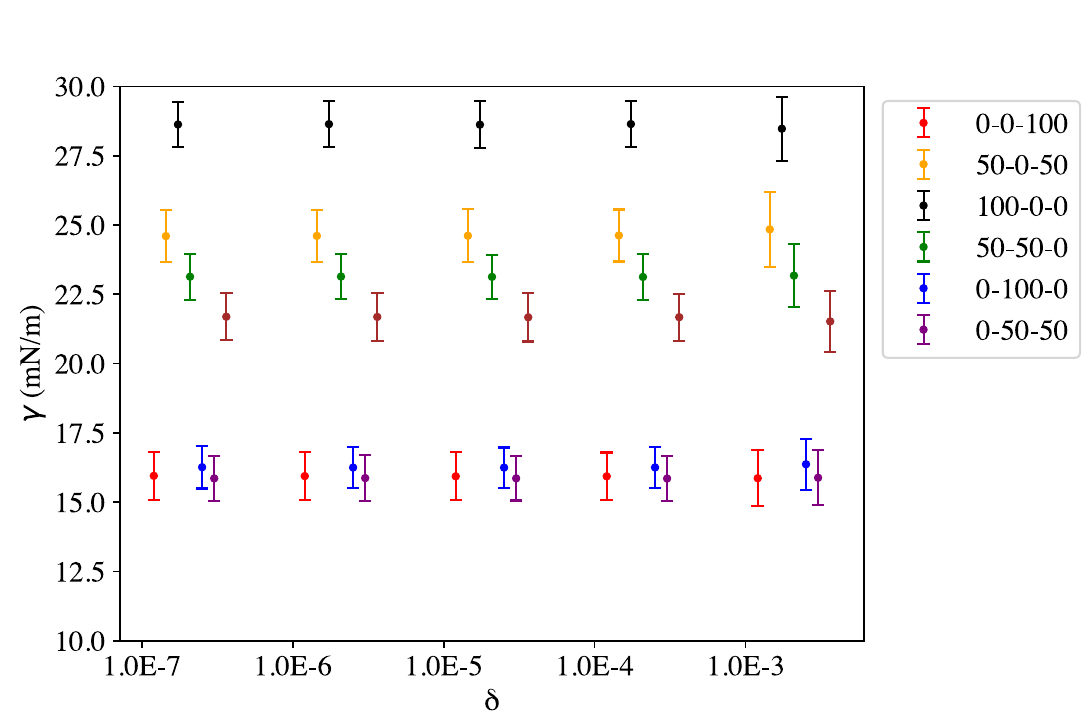}
        \captionsetup{name=Fig, font=footnotesize, singlelinecheck=false}
        \caption{The test area method was applied to a set of mixtures for scaling factors from $10^{-7}$ to $10^{-3}$. Smaller and larger values were found to be affected respectively by noise or discretization errors.}
        \label{fig:test_area}
    \end{minipage}%
    \hfill
    \begin{minipage}{.45\textwidth}
        \centering
            \begin{table}[H]
            \footnotesize
            \begin{adjustbox}{width=\textwidth, center}
            \begin{tabular}{ccc} \hline
            Water composition & $\gamma$ (virial)  & $\gamma$ (test-area) \\ \hline
            W (100\%) & 29.20 $\pm$ 0.42 & 28.63 $\pm$ 0.85 \\ 
            SW (100\%) & 16.03 $\pm$ 0.38 & 16.26 $\pm$ 0.73 \\ 
            TW (100\%) & 16.33 $\pm$ 0.44 & 15.94 $\pm$ 0.84 \\ 
            W-SW-TW (33\%) & 20.71 $\pm$ 0.41 & 21.67 $\pm$ 0.85 \\ \hline
            \end{tabular}
            \end{adjustbox}
            \captionsetup{font=footnotesize, singlelinecheck=false}
            \caption{Comparison between surface tension values (in mN/m) of selected water compositions, computed with the virial and test-area methods. Errors were estimated using the bootstrapping technique \cite{efron1994_bootstrap}.}
            \end{table}
            \label{tab:virial}
    \end{minipage}
\end{figure}

 Simulations were conducted using the LAMMPS \cite{LAMMPS} software (August 2023 stable release) and Martini 3 forcefield parameters for water. Test area for computing the surface tension was performed using the built-in LAMMPS function compute fep/ta (\url{https://docs.lammps.org/compute_fep_ta.html}) 
which is based on the seminal Test-Area paper by Gloor et al. \cite{gloor2005}. The scaling factor $\delta$ used in the perturbation of the \emph{l-v} interface $A_{lv}$ was chosen in an interval such that the resulting surface tension $\left(\frac{\partial F}{\partial A_{lv}}\right)_T$ was found to be indepedent of the perturbation (Fig.~\ref{fig:test_area}). To assess the independence of the results on the specific molecular dynamics engine and computation method, we calculated a few values of surface tension with GROMACS\cite{van2005_gromacs} using the virial method, obtaining reasonable values within a 10\% difference from the test area ones (Tab. I). In Appendix~\ref{appendix:virial} we give a short description of the virial method approach and of the procedure followed to equilibrate the system and compute the surface tension.

\subsection{Virial method}
\label{appendix:virial}

        To build the liquid-vapour system, we used a water box made of beads of all three different sizes (W, SW, TW) in different percentages (100\% Regular water W, 100\% SW, 100\% TW and 33\% for all the three bead types). The water box contained a total of 11000 beads, and was equilibrated in NPT conditions to reach the correct density. Then, we increased the box length along $z$ to create two water-vacuum interfaces (on the $x-y$ plane). The box has volume $V=L_x L_y L_z$, with $L_x=L_y=10$~nm and $L_z=30$~nm. Periodic boundary conditions are enforced in all dimensions.
        \\ According to the virial method, thoroughly explained and derived in ref.\cite{KBmethod}, we computed the surface tension as:
        \begin{equation}
            \gamma = \frac{L_z}{2} (\langle p_N-p_T\rangle)
        \end{equation}
        with $\langle \cdot \rangle$ the ensemble average, $p_N$ and $p_T$ respectively the normal and tangential components of pressure, and $L_z$ the length of the box in the direction perpendicular to the liquid-vacuum interfaces. Referring to the pressure tensor, we have $p_N=p_{zz}$ and $p_T=\frac{1}{2}(p_{xx}+p_{yy})$.
        The components of the pressure ($p_{xx}, p_{yy}, p_{zz}$) result from a $1~\mathrm{\mu s}$ simulation at constant volume. The software used to perform molecular dynamics was GROMACS 2022. We used a leap-frog stochastic dynamics integrator (\textit{sd}), which adds a friction and a noise term to Newton’s equations of motion, to get a better stability of the simulations in presence of vacuum. The typical Martini time step of 20 fs was used and the temperature of the system was set to $T=300$~K.

\newpage

\subsection{Nominal vs. actual equilibrium molar fractions}
\label{appendix:mol_fractions}

\begin{table}[H]
    \centering
    \adjustbox{max width=\columnwidth}{
    \renewcommand{\arraystretch}{0.8}
    \begin{tabular}{c@{\hspace{15pt}}c@{\hspace{15pt}}c@{\hspace{15pt}}c}
       & Initial. liquid bulk & Equil. liquid bulk & Equil. vapour bulk \\ \hline\hline
       & W - SW - TW & W - SW - TW & W - SW - TW \\
0-0-100 &	  0.00 - 0.00 - 1.00 &  0.00 - 0.00 - 1.00 &  0.00 - 0.00 - 1.00 \\
10-0-90 &	  0.10 - 0.00 - 0.90 &  0.11 - 0.00 - 0.89 &  0.00 - 0.00 - 1.00 \\
20-0-80 &	  0.20 - 0.00 - 0.80 &  0.22 - 0.00 - 0.78 &  0.01 - 0.00 - 0.99 \\
30-0-70 &	  0.30 - 0.00 - 0.70 &  0.33 - 0.00 - 0.67 &  0.01 - 0.00 - 0.99 \\
40-0-60 &	  0.40 - 0.00 - 0.60 &  0.43 - 0.00 - 0.57 &  0.02 - 0.00 - 0.98 \\
50-0-50 &	  0.50 - 0.00 - 0.50 &  0.53 - 0.00 - 0.47 &  0.03 - 0.00 - 0.97 \\
60-0-40 &	  0.60 - 0.00 - 0.40 &  0.63 - 0.00 - 0.37 &  0.04 - 0.00 - 0.96 \\
70-0-30 &	  0.70 - 0.00 - 0.30 &  0.73 - 0.00 - 0.27 &  0.06 - 0.00 - 0.94 \\
80-0-20 &	  0.80 - 0.00 - 0.20 &  0.82 - 0.00 - 0.18 &  0.10 - 0.00 - 0.90 \\
90-0-10 &	  0.90 - 0.00 - 0.10 &  0.91 - 0.00 - 0.09 &  0.20 - 0.00 - 0.80 \\
100-0-0 &	  1.00 - 0.00 - 0.00 &  1.00 - 0.00 - 0.00 &  1.00 - 0.00 - 0.00 \\
90-10-0 &	  0.90 - 0.10 - 0.00 &  0.90 - 0.10 - 0.00 &  0.26 - 0.74 - 0.00 \\
80-20-0 &	  0.80 - 0.20 - 0.00 &  0.82 - 0.18 - 0.00 &  0.01 - 0.99 - 0.00 \\
70-30-0 &	  0.70 - 0.30 - 0.00 &  0.72 - 0.28 - 0.00 &  0.10 - 0.90 - 0.00 \\
60-40-0 &	  0.60 - 0.40 - 0.00 &  0.63 - 0.37 - 0.00 &  0.08 - 0.92 - 0.00 \\
50-50-0 &	  0.50 - 0.50 - 0.00 &  0.53 - 0.47 - 0.00 &  0.06 - 0.94 - 0.00 \\
40-60-0 &	  0.40 - 0.60 - 0.00 &  0.43 - 0.57 - 0.00 &  0.00 - 1.00 - 0.00 \\
30-70-0 &	  0.30 - 0.70 - 0.00 &  0.32 - 0.68 - 0.00 &  0.03 - 0.97 - 0.00 \\
20-80-0 &	  0.20 - 0.80 - 0.00 &  0.21 - 0.79 - 0.00 &  0.02 - 0.98 - 0.00 \\
10-90-0 &	  0.10 - 0.90 - 0.00 &  0.11 - 0.89 - 0.00 &  0.00 - 1.00 - 0.00 \\
0-100-0 &	  0.00 - 1.00 - 0.00 &  0.00 - 1.00 - 0.00 &  0.00 - 1.00 - 0.00 \\
0-90-10 &	  0.00 - 0.90 - 0.10 &  0.00 - 0.90 - 0.10 &  0.00 - 0.80 - 0.20 \\
0-80-20 &	  0.00 - 0.80 - 0.20 &  0.00 - 0.81 - 0.19 &  0.00 - 0.63 - 0.37 \\
0-70-30 &	  0.00 - 0.70 - 0.30 &  0.00 - 0.71 - 0.29 &  0.00 - 0.51 - 0.49 \\
0-60-40 &	  0.00 - 0.60 - 0.40 &  0.00 - 0.61 - 0.39 &  0.00 - 0.40 - 0.60 \\
0-50-50 &	  0.00 - 0.50 - 0.50 &  0.00 - 0.51 - 0.49 &  0.00 - 0.31 - 0.69 \\
0-40-60 &	  0.00 - 0.40 - 0.60 &  0.00 - 0.41 - 0.59 &  0.00 - 0.23 - 0.77 \\
0-30-70 &	  0.00 - 0.30 - 0.70 &  0.00 - 0.31 - 0.69 &  0.00 - 0.16 - 0.84 \\
0-20-80 &	  0.00 - 0.20 - 0.80 &  0.00 - 0.21 - 0.79 &  0.00 - 0.10 - 0.90 \\
0-10-90 &	  0.00 - 0.10 - 0.90 &  0.00 - 0.10 - 0.90 &  0.00 - 0.05 - 0.95 \\
33-33-33 &	  0.33 - 0.33 - 0.33 &  0.36 - 0.32 - 0.32 &  0.02 - 0.31 - 0.67\\       
    \end{tabular} }
    \caption{Nominal molar fractions of beads and the corresponding equilibrium compositions computed in the liquid and vapour bulks after equilibration.}
    \label{tab:equilfraction}
\end{table}

\subsection{Finite size effects}
\label{appendix:size_effects}
Finite size effects are known to affect the value of the surface tension and its determination by means of simulations \cite{binder1982}. For instance, on the occasion of the second release of the Martini forcefield, \cite{marrink2007martini} the surface tension of Martini water at $T=293$~K was measured at $\gamma=30$~mN/m  for a small ($400$ CG water beads) and at $\gamma=45$~mN/m for a larger ($1600$ CG water beads) system. This discrepancy could be attributed to finite size effects, and simulations with water boxes containing more than $1600$ beads showed no further deviations from the value of $\gamma=45$~mN/m. Goujon et al.\cite{goujon2018does} showed that the finite size effects resulted in small oscillations of around $2$~mN~m$^{-1}$, in a range of interfacial areas from $4\times 4 ~\mathrm{nm} ^{2}$ to $10\times 10 ~ \mathrm{nm} ^{2}$ for polarizable Martini water. In Tab.~\ref{tab:size_related_effects} we present the values of surface tension that were measured simulating systems of different sizes. These results show how, at the sizes we considered, size-dependent contributions are subleading and smaller than the statistical errors that affect our measurements. This provides justification to our choice to neglect finite-size corrections to the surface tension in our simulation campaign. 

\begin{table}[H]
    \centering
    \begin{tabular}{c@{\hspace{15pt}}c} \hline\hline
        size & $\gamma(mN/m)$  \\ \hline
        50x50x150 & 23.0510 $\pm$ 1.11 \\ 
        75x75x225 & 22.6652 $\pm$ 0.94 \\
        \bf{100x100x300} & \bf{23.1225} $\pm$ 0.81  \\
        125x125x375 & 22.0390 $\pm$ 0.76 \\
        150x150x450 & 22.9766 $\pm$ 0.69 \\
        175x175x525 & 22.4939 $\pm$ 0.63 \\
        200x200x600 & 22.6607 $\pm$ 0.59 \\
    \end{tabular}
    \caption{Surface tension computed for a 50\%-50 \% W-SW mixture at 300 K. Box sizes are expressed in Å.}
    \label{tab:size_related_effects}
\end{table}

\newpage 

\subsection{Tables for surface tension at 300, 310, 320, 330, 340, 350 K}
\label{appendix:T_data}

\begin{table}[H]
\centering
\caption{Surface tension at 300 K}
\begin{minipage}{0.33\textwidth}
\begin{table}[H]
\begin{tabular}{ccc} \hline
W-SW-TW & $\gamma$ (mN/m) \\ \hline
0-0-100   &   15.9363 \\ \hline 
0-0-90   &   18.0842 \\ \hline
20-0-80   &   20.4593 \\ \hline
30-0-70   &   21.8460 \\ \hline
40-0-60   &   23.1425 \\ \hline
50-0-50   &   24.5978 \\ \hline
60-0-40   &   26.0064 \\ \hline
70-0-30   &   26.8657 \\ \hline
80-0-20   &   27.3304 \\ \hline
90-0-10   &   27.5185 \\ \hline
100-0-0   &   28.6303 \\ \hline
90-10-0   &   16.0705 \\ \hline
80-20-0   &   26.0415 \\ \hline
70-30-0   &   24.9969 \\ \hline
60-40-0   &   23.4679 \\ \hline
50-50-0   &   23.1294 \\ \hline 
40-60-0   &   20.8685 \\ \hline
30-70-0   &   20.2579 \\ \hline
20-80-0   &   18.4637 \\ \hline
10-90-0   &   17.3165 \\ \hline 
0-100-0   &   16.2617 \\ \hline
0-90-10   &   16.2080 \\ \hline
\end{tabular}
\end{table}
\end{minipage}
\hspace{-4em}
\begin{minipage}{0.33\textwidth}
\begin{table}[H]
\begin{tabular}{ccc} \hline
W-SW-TW & $\gamma$ (mN/m) \\ \hline
0-80-20   &   16.1049 \\ \hline
0-70-30   &   15.7156 \\ \hline
0-60-40   &   15.8146 \\ \hline
0-50-50   &   15.8649 \\ \hline
0-40-60   &   16.6727 \\ \hline
0-30-70   &   16.0554 \\ \hline
0-20-80   &   15.7607 \\ \hline
0-10-90   &   15.7205 \\ \hline
10-80-10   &   17.1728 \\ \hline
20-70-10   &   18.5443  \\ \hline
30-60-10   &   19.6383 \\ \hline
40-50-10   &   20.9484 \\ \hline
50-40-10   &   23.2600 \\ \hline
60-30-10   &   23.9320 \\ \hline
70-20-10   &   24.8290 \\ \hline
80-10-10   &   26.3655 \\ \hline
10-70-20   &   17.5504 \\ \hline
20-60-20   &   19.5274\\ \hline
30-50-20   &   20.2382 \\ \hline
40-40-20   &   22.0007 \\ \hline
50-30-20   &   23.2668 \\ \hline
60-20-20   &   24.6607 \\ \hline
\end{tabular}
\end{table}
\end{minipage}
\hspace{-4em}
\begin{minipage}{0.33\textwidth}
\begin{table}[H]
\begin{tabular}{ccc} \hline
W-SW-TW & $\gamma$ (mN/m) \\ \hline
70-10-20   &   25.9627 \\ \hline
20-50-30   &   19.6403 \\ \hline
30-40-30   &   20.8014 \\ \hline
40-30-30   &   22.3790 \\ \hline
50-20-30   &   23.7831 \\ \hline
60-10-30   &   25.6338 \\ \hline
10-50-40   &   17.9926 \\ \hline
10-60-30   &   17.7763 \\ \hline
20-40-40   &   19.1824 \\ \hline
30-30-40   &   20.8804 \\ \hline
40-20-40   &   22.7181 \\ \hline
50-10-40   &   24.5715 \\ \hline
10-40-50   &   18.2224 \\ \hline
20-30-50   &   19.9227 \\ \hline
30-20-50   &   21.3612 \\ \hline
40-10-50   &   22.8786. \\ \hline 
10-30-60   &   17.7930. \\ \hline 
20-20-60   &   20.3353 \\ \hline
30-10-60   &   21.4452 \\ \hline
10-20-70   &   17.6263 \\ \hline
10-10-80   &   18.8219  \\ \hline
20-10-70   &   20.3900 \\ \hline
\end{tabular}
\end{table}
\end{minipage}
\begin{minipage}{0.33\textwidth}
\begin{table}[H]
\begin{tabular}{ccc} \hline
W-SW-TW & $\gamma$ (mN/m) \\ \hline
33-33-33   &   21.6750 \\ \hline
\end{tabular}
\end{table}
\end{minipage}
\end{table}

\newpage

\begin{table}[H]
\centering
\caption{Surface tension at 310 K}
\begin{minipage}{0.33\textwidth}
\begin{table}[H]
\begin{tabular}{ccc} \hline
W-SW-TW & $\gamma$ (mN/m) \\ \hline
0-0-100    &    13.3988 \\ \hline
10-0-90    &    15.5888  \\ \hline
20-0-80    &    18.1256  \\ \hline
30-0-70    &     19.9781  \\ \hline
40-0-60    &    20.9349  \\ \hline
50-0-50    &    22.0496  \\ \hline
60-0-40    &    23.5317  \\ \hline
70-0-30    &    24.0071  \\ \hline
80-0-20    &    24.9283   \\ \hline
90-0-10    &    26.0567  \\ \hline
100-0-0    &    26.4074  \\ \hline
90-10-0    &     25.7360 \\ \hline
80-20-0    &    24.0981  \\ \hline
70-30-0    &    23.1041  \\ \hline
60-40-0    &    21.9575  \\ \hline
50-50-0    &    20.3637  \\ \hline
40-60-0    &    19.0659  \\ \hline
30-70-0    &    18.2886  \\ \hline
20-80-0    &    16.9645  \\ \hline
10-90-0    &    15.5139  \\ \hline
0-100-0    &    14.5640 \\ \hline
0-90-10    &    14.8485  \\ \hline
\end{tabular}
\end{table}
\end{minipage}
\hspace{-4em}
\begin{minipage}{0.33\textwidth}
\begin{table}[H]
\begin{tabular}{ccc} \hline
W-SW-TW & $\gamma$ (mN/m) \\ \hline
0-80-20    &    14.1013  \\ \hline
0-70-30    &    14.1336  \\ \hline
0-60-40    &    14.2678 \\ \hline 
0-50-50    &    13.8821  \\ \hline
0-40-60    &    13.8001  \\ \hline
0-30-70    &    13.8465  \\ \hline
0-20-80    &    13.7562  \\ \hline
0-10-90    &    13.9346  \\ \hline
10-80-10   &     15.4383 \\ \hline
20-70-10   &     16.6945 \\ \hline
30-60-10   &     17.9585 \\ \hline
40-50-10   &     19.4347 \\ \hline
50-40-10   &     20.7948 \\ \hline
60-30-10   &     22.5291 \\ \hline
70-20-10   &     23.0983 \\ \hline
80-10-10   &     24.7162 \\ \hline
10-70-20   &     15.6988 \\ \hline
20-60-20   &     17.2618 \\ \hline 
30-50-20   &     18.2458 \\ \hline
40-40-20   &     20.5221 \\ \hline
50-30-20   &     21.7645 \\ \hline
60-20-20   &     22.4319 \\ \hline
\end{tabular}
\end{table}
\end{minipage}
\hspace{-4em}
\begin{minipage}{0.33\textwidth}
\begin{table}[H]
\begin{tabular}{ccc} \hline
W-SW-TW & $\gamma$ (mN/m) \\ \hline
70-10-20   &     24.0188 \\ \hline
10-60-30   &     17.7928 \\ \hline
20-50-30   &     16.9552 \\ \hline
30-40-30   &     19.3403 \\ \hline
40-30-30   &     19.8419 \\ \hline
50-20-30   &     20.9698 \\ \hline
60-10-30   &     23.2397 \\ \hline
10-50-40   &     15.8869  \\ \hline
20-40-40   &     17.4526 \\ \hline
30-30-40   &     17.8074 \\ \hline 
40-20-40   &     19.7227 \\ \hline
50-10-40   &     21.7350 \\ \hline
10-40-50   &     15.6562 \\ \hline
20-30-50   &     16.4740 \\ \hline
30-20-50   &     18.8624 \\ \hline
40-10-50   &     20.5618  \\ \hline
10-30-60   &     15.0858 \\ \hline 
20-20-60   &     16.8869 \\ \hline
30-10-60   &     18.6637 \\ \hline
10-20-70   &     15.7208 \\ \hline
10-10-80   &     15.1017 \\ \hline
20-10-70   &     17.3744 \\ \hline
\end{tabular}
\end{table}
\end{minipage}
\begin{minipage}{0.33\textwidth}
\begin{table}[H]
\begin{tabular}{ccc} \hline
W-SW-TW & $\gamma$ (mN/m) \\ \hline
33-33-33   &     19.1775 \\ \hline
\end{tabular}
\end{table}
\end{minipage}
\end{table}

\newpage

\begin{table}[H]
\centering
\caption{Surface tension at 320 K}
\begin{minipage}{0.33\textwidth}
\begin{table}[H]
\begin{tabular}{ccc} \hline
W-SW-TW & $\gamma$ (mN/m) \\ \hline
0-0-100   &     11.9858 \\ \hline
10-0-90   &     14.2450 \\ \hline
20-0-80   &     16.0612 \\ \hline
30-0-70   &     17.6587 \\ \hline
40-0-60   &     19.1641 \\ \hline
50-0-50   &     21.0498 \\ \hline
60-0-40   &     21.8148 \\ \hline
70-0-30   &     22.7607 \\ \hline
80-0-20   &     23.9085 \\ \hline
90-0-10   &     23.9056 \\ \hline
100-0-0   &     25.5783 \\ \hline
90-10-0   &     24.1315 \\ \hline
80-20-0   &     23.0314 \\ \hline
70-30-0   &     21.8448 \\ \hline
60-40-0   &     20.2010 \\ \hline
50-50-0   &     18.8144 \\ \hline
40-60-0   &     17.6205  \\ \hline
30-70-0   &     16.9784 \\ \hline
20-80-0   &     15.2799 \\ \hline
10-90-0   &     13.8528 \\ \hline
0-100-0   &     12.9435  \\ \hline
0-90-10   &     12.8348 \\ \hline
\end{tabular}
\end{table}
\end{minipage}
\hspace{-4em}
\begin{minipage}{0.33\textwidth}
\begin{table}[H]
\begin{tabular}{ccc} \hline
W-SW-TW & $\gamma$ (mN/m) \\ \hline
0-80-20    &    12.3797 \\ \hline
0-70-30    &    12.0856  \\ \hline
0-60-40    &    12.3875  \\ \hline
0-50-50    &    12.2719 \\ \hline
0-40-60    &    12.2286  \\ \hline
0-30-70    &    12.3891   \\ \hline
0-20-80    &    11.5295  \\ \hline
0-10-90    &    12.0763  \\ \hline
10-80-10   &     13.9628 \\ \hline
20-70-10   &     15.5117 \\ \hline
30-60-10   &     16.6195 \\ \hline
40-50-10   &     17.8992 \\ \hline
50-40-10   &     19.1664 \\ \hline
60-30-10   &     20.5988 \\ \hline
70-20-10   &     22.1621 \\ \hline
80-10-10   &     23.6908 \\ \hline
10-70-20   &     14.1110 \\ \hline
20-60-20   &     15.6490 \\ \hline
30-50-20   &     16.7715 \\ \hline
40-40-20   &     18.9002 \\ \hline
50-30-20   &     19.8136 \\ \hline
60-20-20   &     20.8337 \\ \hline
\end{tabular}
\end{table}
\end{minipage}
\hspace{-4em}
\begin{minipage}{0.33\textwidth}
\begin{table}[H]
\begin{tabular}{ccc} \hline
W-SW-TW & $\gamma$ (mN/m) \\ \hline
70-10-20    &    22.1545 \\ \hline
10-60-30    &    14.0563 \\ \hline
20-50-30    &    15.3989 \\ \hline
30-40-30    &    17.3062 \\ \hline
40-30-30    &    18.4862 \\ \hline
50-20-30    &    19.8837 \\ \hline
60-10-30    &    21.6558 \\ \hline
10-50-40    &    14.0017 \\ \hline
20-40-40    &    15.4151 \\ \hline
30-30-40    &    17.5086 \\ \hline
40-20-40    &    18.4360 \\ \hline
50-10-40    &    19.6299 \\ \hline
10-40-50    &    14.1695 \\ \hline
20-30-50    &    15.1434 \\ \hline
30-20-50    &    17.6740 \\ \hline
40-10-50    &    18.8414 \\ \hline
10-30-60    &    14.0823 \\ \hline
20-20-60    &    15.8088 \\ \hline
30-10-60    &    17.6482 \\ \hline
10-20-70    &    14.2570 \\ \hline
10-10-80    &    14.0729 \\ \hline
20-10-70    &    16.2916 \\ \hline
\end{tabular}
\end{table}
\end{minipage}
\begin{minipage}{0.33\textwidth}
\begin{table}[H]
\begin{tabular}{ccc} \hline
W-SW-TW & $\gamma$ (mN/m) \\ \hline
33-33-33    &    17.8992 \\ \hline
\end{tabular}
\end{table}
\end{minipage}
\end{table}

\newpage

\begin{table}[H]
\centering
\caption{Surface tension at 330 K}
\begin{minipage}{0.33\textwidth}
\begin{table}[H]
\begin{tabular}{ccc} \hline
W-SW-TW & $\gamma$ (mN/m) \\ \hline
0-0-100   &     9.6145 \\ \hline
10-0-90   &     12.2497 \\ \hline
20-0-80   &     15.2850 \\ \hline
30-0-70   &     16.1854 \\ \hline
40-0-60   &     18.0553 \\ \hline
50-0-50   &     18.5542 \\ \hline
60-0-40   &     19.9809 \\ \hline
70-0-30   &     22.1106 \\ \hline
80-0-20   &     22.7028 \\ \hline
90-0-10   &     23.7416 \\ \hline
100-0-0   &     24.0764 \\ \hline
90-10-0   &     22.8704 \\ \hline
80-20-0   &     21.4402 \\ \hline
70-30-0   &     21.0140 \\ \hline
60-40-0   &     19.2594 \\ \hline
50-50-0   &     17.9414 \\ \hline
40-60-0   &     16.9106 \\ \hline
30-70-0   &     15.4339  \\ \hline
20-80-0   &     13.9250 \\ \hline
10-90-0   &     13.0597 \\ \hline
0-100-0   &     10.9871 \\ \hline
0-90-10   &     11.4714 \\ \hline
\end{tabular}
\end{table}
\end{minipage}
\hspace{-4em}
\begin{minipage}{0.33\textwidth}
\begin{table}[H]
\begin{tabular}{ccc} \hline
W-SW-TW & $\gamma$ (mN/m) \\ \hline
0-80-20   &     11.0054 \\ \hline
0-70-30   &     11.1831  \\ \hline
0-60-40   &     11.3885 \\ \hline
0-50-50   &     10.5707  \\ \hline
0-40-60   &     10.6349  \\ \hline
0-30-70   &     10.4001 \\ \hline
0-20-80   &     10.0629  \\ \hline
0-10-90   &      9.7006  \\ \hline
10-80-10  &      12.4634 \\ \hline
20-70-10  &      14.1272 \\ \hline
30-60-10  &      15.9283 \\ \hline
40-50-10  &      16.6100 \\ \hline
50-40-10  &      17.8660 \\ \hline
60-30-10  &      19.2599 \\ \hline
70-20-10  &      20.6867 \\ \hline
80-10-10  &      22.3675 \\ \hline
10-70-20  &      12.7586 \\ \hline
20-60-20  &      13.6787 \\ \hline
30-50-20  &      15.3096 \\ \hline
40-40-20  &      17.3067 \\ \hline
50-30-20  &      18.6086 \\ \hline
60-20-20  &      20.150 \\ \hline
\end{tabular}
\end{table}
\end{minipage}
\hspace{-4em}
\begin{minipage}{0.33\textwidth}
\begin{table}[H]
\begin{tabular}{ccc} \hline
W-SW-TW & $\gamma$ (mN/m) \\ \hline
70-10-20   &     21.3296 \\ \hline
10-60-30   &     12.4274 \\ \hline
20-50-30   &     14.5080 \\ \hline
30-40-30   &     15.9695 \\ \hline
40-30-30   &     16.7805 \\ \hline
50-20-30   &     18.6575 \\ \hline
60-10-30   &     20.0953 \\ \hline
10-50-40   &     12.1074 \\ \hline
20-40-40   &     13.5981 \\ \hline
30-30-40   &     15.6697 \\ \hline
40-20-40   &     16.8532 \\ \hline
50-10-40   &     18.9061 \\ \hline
10-40-50   &     12.8288 \\ \hline
20-30-50   &     14.3825 \\ \hline
30-20-50   &     15.5843 \\ \hline
40-10-50   &     17.5997 \\ \hline
10-30-60   &     12.7043 \\ \hline
20-20-60   &     14.0579 \\ \hline
30-10-60   &     15.6843 \\ \hline
10-20-70   &     12.1690 \\ \hline
10-10-80   &     12.2206 \\ \hline
20-10-70   &     14.8194 \\ \hline
\end{tabular}
\end{table}
\end{minipage}
\hspace{-4em}
\begin{minipage}{0.33\textwidth}
\begin{table}[H]
\begin{tabular}{ccc} \hline
W-SW-TW & $\gamma$ (mN/m) \\ \hline
33-33-33    &    16.5811 \\ \hline
\end{tabular}
\end{table}
\end{minipage}
\hspace{-4em}
\end{table}

\newpage

\begin{table}[H]
\centering
\caption{Surface tension at 340 K}
\begin{minipage}{0.33\textwidth}
\begin{table}[H]
\begin{tabular}{ccc} \hline
W-SW-TW & $\gamma$ (mN/m) \\ \hline
0-0-100   &      8.5068 \\ \hline
10-0-90   &     10.9574 \\ \hline
20-0-80   &     13.2499  \\ \hline
30-0-70   &     15.1668 \\ \hline
40-0-60   &     16.6604 \\ \hline
50-0-50   &     17.2923 \\ \hline
60-0-40   &     18.9715 \\ \hline
70-0-30   &     20.2990 \\ \hline
80-0-20   &     21.4226 \\ \hline
90-0-10   &     22.1936 \\ \hline
100-0-0   &     23.0231 \\ \hline
90-10-0   &     21.1682 \\ \hline
80-20-0   &     20.4357 \\ \hline
70-30-0   &     19.3086 \\ \hline
60-40-0   &     17.2900 \\ \hline
50-50-0   &     16.3253 \\ \hline
40-60-0   &     15.2872 \\ \hline
30-70-0   &     13.9717 \\ \hline
20-80-0   &     12.5660 \\ \hline
10-90-0   &     11.4739 \\ \hline
0-100-0   &      9.9588 \\ \hline
0-90-10   &     10.0636 \\ \hline
\end{tabular}
\end{table}
\end{minipage}
\hspace{-4em}
\begin{minipage}{0.33\textwidth}
\begin{table}[H]
\begin{tabular}{ccc} \hline
W-SW-TW & $\gamma$ (mN/m) \\ \hline
0-80-20    &     9.6906 \\ \hline
0-70-30    &     9.8204  \\ \hline 
0-60-40    &     9.3294 \\ \hline 
0-50-50    &     9.0206 \\ \hline 
0-40-60    &     9.4200 \\ \hline 
0-30-70    &     8.8478  \\ \hline 
0-20-80    &     8.5659 \\ \hline 
0-10-90    &     8.5763  \\ \hline
10-80-10   &     11.1830 \\ \hline
20-70-10   &     12.4671 \\ \hline
30-60-10   &     14.3534 \\ \hline
40-50-10   &     15.4468 \\ \hline
50-40-10   &     16.9806 \\ \hline
60-30-10   &     17.8265 \\ \hline
70-20-10   &     19.8219 \\ \hline
80-10-10   &     20.9623 \\ \hline
10-70-20   &     11.4233 \\ \hline
20-60-20   &     12.4955 \\ \hline
30-50-20   &     14.2317 \\ \hline
40-40-20   &     15.7418 \\ \hline
50-30-20   &     17.1410 \\ \hline
60-20-20   &     18.1403 \\ \hline
\end{tabular}
\end{table}
\end{minipage}
\hspace{-4em}
\begin{minipage}{0.33\textwidth}
\begin{table}[H]
\begin{tabular}{ccc} \hline
W-SW-TW & $\gamma$ (mN/m) \\ \hline
70-10-20    &    19.9288 \\ \hline
10-60-30    &    11.4583 \\ \hline
20-50-30    &    12.7187 \\ \hline
30-40-30    &    14.3239 \\ \hline
40-30-30    &    15.5252 \\ \hline
50-20-30    &    17.8733 \\ \hline
60-10-30    &    18.6384 \\ \hline
10-50-40    &    11.4985 \\ \hline
20-40-40    &    12.5006 \\ \hline
30-30-40    &    14.2840 \\ \hline
40-20-40    &    15.6636 \\ \hline
50-10-40    &    17.7105 \\ \hline
10-40-50    &    10.8313 \\ \hline
20-30-50    &    12.3688 \\ \hline
30-20-50    &    14.3910 \\ \hline
40-10-50    &    16.4091 \\ \hline
10-30-60    &    11.0321 \\ \hline
20-20-60    &    12.5438 \\ \hline
30-10-60    &    15.0240 \\ \hline
10-20-70    &    10.5772 \\ \hline
10-10-80    &    11.3040 \\ \hline
20-10-70    &    13.0929 \\ \hline
\end{tabular}
\end{table}
\end{minipage}
\hspace{-4em}
\begin{minipage}{0.33\textwidth}
\begin{table}[H]
\begin{tabular}{ccc} \hline
W-SW-TW & $\gamma$ (mN/m) \\ \hline
33-33-33    &    15.2645 \\ \hline
\end{tabular}
\end{table}
\end{minipage}
\hspace{-4em}
\end{table}

\newpage

\begin{table}[H]
\centering
\caption{Surface tension at 350 K}
\begin{minipage}{0.33\textwidth}
\begin{table}[H]
\begin{tabular}{ccc} \hline
W-SW-TW & $\gamma$ (mN/m) \\ \hline
0-0-100   &      9.3531  \\ \hline 
10-0-90   &      9.3409 \\ \hline 
20-0-80   &     11.4056 \\ \hline 
30-0-70   &     13.6435 \\ \hline 
40-0-60   &     15.1460 \\ \hline 
50-0-50   &     17.1308 \\ \hline 
60-0-40   &     17.3439 \\ \hline 
70-0-30   &     18.7455 \\ \hline 
80-0-20   &     20.1233 \\ \hline 
90-0-10   &     20.8065 \\ \hline
100-0-0   &     21.8678 \\ \hline
90-10-0   &     20.0827 \\ \hline 
80-20-0   &     18.8507 \\ \hline 
70-30-0   &     18.1706 \\ \hline 
60-40-0   &     16.8790 \\ \hline 
50-50-0   &     15.7878 \\ \hline 
40-60-0   &     13.9569 \\ \hline 
30-70-0   &     12.3749 \\ \hline 
20-80-0   &     11.3996 \\ \hline 
10-90-0   &     10.2897 \\ \hline
0-100-0   &      8.8620 \\ \hline
0-90-10   &      8.0721 \\ \hline
\end{tabular}
\end{table}
\end{minipage}
\hspace{-4em}
\begin{minipage}{0.33\textwidth}
\begin{table}[H]
\begin{tabular}{ccc} \hline
W-SW-TW & $\gamma$ (mN/m) \\ \hline
0-80-20    &     8.4850 \\ \hline
0-70-30    &     8.3294 \\ \hline
0-60-40    &     8.4112 \\ \hline
0-50-50    &     7.9456 \\ \hline
0-40-60    &     7.1696 \\ \hline
0-30-70    &     7.1061 \\ \hline
0-20-80    &     6.8491 \\ \hline
0-10-90    &     7.4109 \\ \hline
10-80-10   &     9.8160 \\ \hline
20-70-10   &     11.1231 \\ \hline
30-60-10   &     13.0611 \\ \hline
40-50-10   &     14.4537 \\ \hline
50-40-10   &     15.6753 \\ \hline
60-30-10   &     16.8482 \\ \hline
70-20-10   &     18.6148 \\ \hline
80-10-10   &     19.8008 \\ \hline
10-70-20   &      9.9038 \\ \hline
20-60-20   &     11.2153 \\ \hline
30-50-20   &     12.9359 \\ \hline
40-40-20   &     14.6221 \\ \hline
50-30-20   &     16.1886 \\ \hline
60-20-20   &     17.0305 \\ \hline
\end{tabular}
\end{table}
\end{minipage}
\hspace{-4em}
\begin{minipage}{0.33\textwidth}
\begin{table}[H]
\begin{tabular}{ccc} \hline
W-SW-TW & $\gamma$ (mN/m) \\ \hline
70-10-20    &    18.1390 \\ \hline
10-60-30    &     9.7988 \\ \hline 
20-50-30    &    11.0921 \\ \hline
30-40-30    &    12.7802 \\ \hline
40-30-30    &    14.5947 \\ \hline
50-20-30    &    15.9211 \\ \hline
60-10-30    &    17.5340 \\ \hline
10-50-40    &     9.6756 \\ \hline
20-40-40    &    10.9399 \\ \hline
30-30-40    &    13.2378 \\ \hline
40-20-40    &    14.7248 \\ \hline
50-10-40    &    16.1894 \\ \hline
10-40-50    &     9.1488 \\ \hline
20-30-50    &    11.0986 \\ \hline
30-20-50    &    13.2447 \\ \hline
40-10-50    &    15.1786 \\ \hline
10-30-60    &     9.2331 \\ \hline
20-20-60    &    11.4975 \\ \hline
30-10-60    &    13.7813 \\ \hline
10-20-70    &     8.8700 \\ \hline
10-10-80    &     9.2875 \\ \hline
20-10-70    &    11.3189 \\ \hline
\end{tabular}
\end{table}
\end{minipage}
\hspace{-4em}
\begin{minipage}{0.33\textwidth}
\begin{table}[H]
\begin{tabular}{ccc} \hline
W-SW-TW & $\gamma$ (mN/m) \\ \hline
33-33-33    &    13.7101 \\ \hline
\end{tabular}
\end{table}
\end{minipage}
\hspace{-4em}
\end{table}

\begin{acknowledgments}

We acknowledge financial support under the National Recovery and Resilience Plan (NRRP), Mission 4, Component 2, Investment 1.1, Call for tender No. 104 published on 2.2.2022 by the Italian Ministry of University and Research (MUR), funded by the European Union – NextGenerationEU– Project Title SoftNanoPores – CUP D53D23002240006 - Grant Assignment Decree No. 957 adopted on 30/06/2023 by the Italian Ministry of Ministry of University and Research (MUR). LI acknowledges financial support under the National Recovery and Resilience Plan (NRRP), CUP B53C23002020006. SC acknowledges funding from the SUNSHINE project (H2020, grant number 952924). We acknowledge the EuroHPC Joint Undertaking for awarding access to the EuroHPC supercomputer LUMI, hosted by CSC (Finland) and the LUMI consortium through a EuroHPC Regular Access call.

\end{acknowledgments}

\section*{Data Availability Statement}
Interfacial tension and bulk density measurements at all 300, 310, 320, 330, 340, 350 K as well as values of $T_c$ and temperature scaling coefficients  used in ~\ref{eqn:gamma_scaling_law}, ~\ref{eqn:density_scaling_law} for all mixtures computed in this study are made available as open data in a repository hosted by Zenodo. The repository is accessible through the Digital Object Identifier \emph{doi:10.5281/zenodo.11369509} .

\nocite{*}
\bibliography{bibliography}

\end{document}